\documentstyle[preprint,aps]{revtex}
\tighten
\draft
\widetext
\input epsf
\preprint{CLNS 98/1550, HUTP-97/A092, NUB 3173}
\bigskip
\bigskip
\begin{document}
\title{Type IIB Orientifolds with NS-NS Antisymmetric
Tensor Backgrounds}
\medskip
\author{Zurab Kakushadze$^{1,2}$\footnote{E-mail: 
zurab@string.harvard.edu},
Gary Shiu$^3$\footnote{E-mail: shiu@mail.lns.cornell.edu} and S.-H. Henry
Tye$^3$\footnote{E-mail: tye@mail.lns.cornell.edu}}
\bigskip
\address{$^1$Lyman Laboratory of Physics, Harvard University, Cambridge, 
MA 02138\\
$^2$Department of Physics, Northeastern University, Boston, MA 02115\\
$^3$Newman Laboratory of Nuclear Studies, Cornell University,
Ithaca, NY 14853}
\date{March 17, 1998}
\bigskip
\medskip
\maketitle

\begin{abstract}
{}We consider six dimensional ${\cal N}=1$ space-time supersymmetric
Type IIB orientifolds with
non-zero untwisted NS-NS sector $B$-field. 
The $B$-field is quantized due to the requirement that 
the Type IIB spectrum be left-right symmetric. 
The presence of the $B$-field
results in rank reduction of both $99$ and $55$ open string sector 
gauge groups.
We point out that in some of the 
models with non-zero $B$-field 
there are extra tensor multiplets in the ${\bf Z}_2$ twisted 
closed string sector, and we explain their origin 
in a simple example. Also, the 59 open string sector states come
with a multiplicity that depends on the $B$-field.
These two facts are in accord with anomaly cancellation
requirements. We point out relations between various orientifolds with 
and without the $B$-field, and also discuss the F-theory duals of these 
models.
\end{abstract}

\section{Introduction}

{}In recent years a lot of progress has been made in understanding
the web of dualities between different string theories in various 
dimensions. Among the
five consistent string theories in ten dimensions, Type I string theory
(which is a theory of unoriented open plus closed strings) remains 
largely unexplored, especially upon compactification with a non-trivial 
background. 
This is in part due to lack of modular invariance 
which is necessary for perturbative consistency of oriented closed
string theories.

{}Various unoriented open plus closed string vacua have been constructed
using orientifold techniques 
\cite{PS,GP,6Dorientifold,6Dduality,BL,Sagnotti,ZK,KS,Zw}. 
Some of these vacua exhibit novel
features that
are inaccessible perturbatively in other string theories.
For instance, upon compactification to six dimensions it is
possible to construct ${\cal N}=1$ orientifolds with multiple tensor
multiplets \cite{PS,6Dorientifold,6Dduality}. 
Similarly, enhanced gauge symmetries due to 
small instantons, which cannot
be described within 
conformal field theory in Heterotic compactifications,
have a perturbative description in terms of coincident D5-branes in the 
Type I language.
Using the conjectured Type I-Heterotic duality \cite{PW}, 
one can understand 
these non-perturbative Heterotic phenomena as perturbative
effects on the Type I side \cite{ZK}.
In fact, this approach has been taken in constructing non-perturbative
chiral Heterotic string vacua in four dimensions \cite{KS}.

{}In this paper we will focus on six\footnote{Four dimensional
${\cal N}=1$ supersymmetric orientifolds will be discussed in \cite{KST}.} 
dimensional ${\cal N}=1$ 
Type IIB orientifolds with constant non-zero NS--NS antisymmetric 
tensor backgrounds. Toroidal compactifications of Type I strings
with non-zero $B$-field have been studied in
\cite{Bij} (and also recently in \cite{bianchi,toroidal}). There it was 
shown that
the rank of the Chan-Paton gauge group is reduced by a factor of
$2^{b/2}$ where $b \in 2 {\bf Z}$ 
is the rank of the matrix $B_{ij}$ corresponding to the compactified 
directions.
In section II we extend these arguments to orbifold compactifications and show
that when there are D9- and/or D5-branes, both the $99$ and $55$ 
gauge group ranks are reduced by a factor of $2^{b/2}$. 
We point out that in some of the 
models with non-zero $B$-field  
there are extra tensor multiplets that arise in the ${\bf Z}_2$ twisted 
closed string sector, and we explain their origin in section III
in a simple example. Also, the 59 open string sector states come
with a multiplicity that depends on the $B$-field.
Appearance of extra tensor multiplets (as well as a non-trivial multiplicity 
of states in the 59 open string sector) is
in accord with anomaly cancellation. 
Thus, cancellation of the ${\mbox{Tr}}(R^4)$ 
gravitational anomalies requires 
\cite{anomalies}:
\begin{equation}
n_H -n_V = 244 - 29 n_T ~,
\end{equation}
where $n_H$, $n_V$ and $n_T +1$ are the numbers of hypermultiplets,
vector multiplets and tensor multiplets, respectively. In the presence of 
the $B$-field, the rank of the gauge group is reduced and hence
the numbers of vector multiplets and hypermultiplets are reduced accordingly.
Unless the difference $n_H -n_V$ is unchanged, this implies that
the number of tensor multiplets
depends upon the quantized value of the $B$-field. 
We will show that some of the fixed points in the ${\bf Z}_2$ twisted
closed string sector become odd  
under the orientifold projection once we turn on the $B$-field.
As a result, tensor multiplets
(instead of hypermultiplets) are kept at those fixed points. 
This gives precisely the correct number of tensor multiplets to
cancel the anomalies.

{}In section IV we point out relations between various orientifolds with 
and without the $B$-field, and also discuss the F-theory duals of these 
models. In particular, some of the models with the $B$-field are on the
same moduli as some of the orientifolds without the $B$-field.
 
\section{Quantized $B$-field and Rank Reduction}\label{ReducedRank}

{}Type IIB orientifolds are generalized orbifolds that involve 
the world-sheet parity reversal $\Omega$ along
with other geometric symmetries of the theory. Orientifolding
results in a theory with unoriented closed strings. In order to cancel
space-time anomalies, the massless tadpoles must be cancelled.
Thus, one generically has to introduce open strings that start and end 
on D-branes.
The global Chan-Paton charges associated with the D-branes give
rise to gauge symmetry in space-time.

{}There are two massless antisymmetric tensor fields in the perturbative 
spectrum of Type IIB theory: one coming from the NS-NS sector, and another 
coming from the R-R sector. Under the world-sheet parity reversal, the NS-NS 
two-form 
$B_{\mu \nu}$ is projected out, while the R-R two-form 
$B_{\mu \nu}^{\prime}$ is kept.
Although the fluctuations of $B_{ij}$ (the components of $B_{\mu \nu}$
in the compactified dimensions) are projected out of the
perturbative unoriented closed string spectrum, a quantized
vacuum expectation value of $B_{ij}$ is allowed. To see this,
consider the left- and right-moving momenta in the
$d$ dimensions compactified on a torus $T^{d}$:
\begin{equation}\label{momenta}
P_{L,R}=\tilde{e}^i (m_i - B_{ij}n^j)  \pm  e_i {n^i \over 2}~,
\end{equation}
where $m_i$ and $n^i$ are integers, $e_i$ are constant vielbeins
such that $e_i \cdot e_j=G_{ij}$ is the constant background metric on 
$T^{d}$, and $e_i \cdot \tilde{e}^{j}={\delta_i}^j$. Note that
the components of $B_{ij}$ are defined up to a shift $B_{ij} \rightarrow
B_{ij} + 1$ (which can be absorbed by redefining $m_i$). With this 
normalization, only the values $B_{ij}=0$ and $1/2$ are invariant under 
$\Omega$, hence quantization of $B_{ij}$.

{}This quantized $B$-field has non-trivial consequences on the
open string spectrum. The effect of non-zero $B$-field 
in toroidal compactifications of Type I string theory
has been studied in \cite{Bij}.
Recall that open strings and D-branes are 
introduced to cancel the 
massless tadpoles coming from the one-loop Klein bottle amplitude
(or, equivalently, the tree channel amplitude for a cylinder with two
cross-caps). In the presence of unoriented open strings there are 
two other
one-loop diagrams which contribute to the tadpoles: the M{\"o}bius strip
and annulus amplitudes. The massless tadpoles are divergences  
in the tree channel due to the exchange of R-R closed string states
between D$p$-branes and/or
orientifold planes both of which carry R-R charges and are sources
for the corresponding $(p+1)$-form potential.
In the case where there are both D9- and D5-branes,
there are three types of massless tadpoles due to the untwisted 
R-R 10-form, untwisted R-R 6-form and twisted R-R 6-form
potentials, respectively.

{}To be specific, let us consider orientifolds of Type IIB on the
orbifold limits of K3: $T^4/{\bf Z}_N$ ($N=2,3,4,6$)\footnote
{Orientifolds of Type IIB on smooth K3 with non-zero $B$-field
have been studied in \cite{SS}. The orbifold cases we are discussing 
here have extra tensor multiplets and qualitatively differ from the 
smooth K3 cases corresponding to Type I compactifications with only one
tensor multiplet.}. Let $g$ be the generator 
of ${\bf Z}_N$. The action of the orbifold on the Chan-Paton factors is described
by matrices $\gamma_{k,9}$ and $\gamma_{k,5}$ corresponding to the action of
the $g^k$ ($k=0,\dots,N-1$) element of ${\bf Z}_N$ on the D9- and D5-branes, 
respectively. Note that ${\mbox{Tr}}(\gamma_{0,9})=n_9$ and 
${\mbox{Tr}}(\gamma_{0,5})=n_5$ are the numbers of D9- and D5-branes,
respectively.

{}First, recall the known results for the case with no 
$B$-field \cite{GP,6Dorientifold}. 
The tadpoles for the untwisted R-R 10-form potential
are proportional to
\begin{equation}
({\mbox {Tr}} (\gamma_{0,9}))^2 - 64 {\mbox {Tr}} (\gamma_{0,9}) + 32^2~,
\end{equation}
and the tadpole cancellation requires presence of $n_9=32$ D9-branes.

{}For even $N$ the tadpoles for the untwisted R-R 6-form potential are
proportional to
\begin{equation}
({\mbox{Tr}} (\gamma_{0,5}))^2 - 64 {\mbox{Tr}} (\gamma_{0,5}) + 32^2~,
\end{equation}
and the tadpole cancellation implies that there are
$n_5=32$ D5-branes. This could also be seen from T-duality between D9- and
D5-branes. For odd $N$ there are no tadpoles for the untwisted R-R 6-form 
potential, and there are no D5-branes.

{}Finally, the tadpole cancellation for the twisted R-R 6-form potential
constrains the action of the twists on the Chan-Paton factors, that is,
the Chan-Paton matrices $\gamma_{k,9}$ and $\gamma_{k,5}$ 
($k=1,\dots,N-1$).
Since the twisted closed string states propagating in the tree channel 
do not have
momenta or windings, the twisted tadpoles will not be affected by
the $B$-field (the effect of which is to shift the left- plus right-moving 
momentum lattice). 
For this reason, we will not need the
explicit expressions for the twisted tadpoles here.

{}Let us now see what happens when we turn on the $B$-field.
Consider the vacuum amplitude for the open string stretched between
two D9-branes. The corresponding graph is an annulus whose
boundaries lie on D9-branes. The open strings stretched between 
D9-branes satisfy Neumann boundary conditions. This implies that in 
the tree-channel
(the corresponding amplitude being a cylinder with closed strings 
propagating between the D9-branes) the closed strings must satisfy the 
condition of ``no momentum flow''
through the boundaries (that is, D9-branes):
\begin{equation}
P_L + P_R = 0~.
\end{equation}
This implies the following constraints on the momenta and windings of the
closed strings propagating along the cylinder:
\begin{equation} \label{constraint}
m_i - B_{ij} n^j = 0~. 
\end{equation}
For $B_{ij}=0$ this simply states that there are no momentum states 
propagating between the D9-branes, but the winding number 
$n^i$ is arbitrary. Let us now consider the cases where some of the 
elements $B_{ij} = 1/2$, so that the rank $b$ of the $B_{ij}$ matrix is 
non-zero. (Note that since $B_{ij}$ is an antisymmetric matrix, its rank
$b$ is always even. For compactifications on orbifold limits of K3 the allowed 
values of
$b$ are $b=0,2,4$.) For simplicity we will assume that 
$T^4=T^2\otimes T^2$ (so that $G_{ij}$ has a $2\times 2$ block-diagonal form),
and the $B_{ij}$ matrix has
a $2\times 2$ block-diagonal form as well. Thus, if 
$B_{ij}=0$, then closed string states with
both odd and even windings $n^j$ contribute into the cylinder 
amplitude. On the other hand, if $B_{ij}=1/2$, then only closed string states with
even windings $n^j$ contribute.  
To see how the cylinder amplitude is modified, let us
extract the piece that depends on $P_L$ and $P_R$ (here we are suppressing 
the contributions from the corresponding oscillators as well as all the other world-sheet 
degrees of freedom for they are not affected by the $B$-field):
\begin{equation}\label{cylinder}
{\cal C} \sim k  \sum_{\bf{n}} e^{- (\pi t/4) G_{ij} n^i n^j}~, 
\end{equation}
where $n^j \in {\bf Z}$ if $B_{ij}=0$, and $n^j \in 2 {\bf Z}$ 
if $B_{ij}=1/2$. Here $t$ is a real modulus (related to the length of the cylinder),
and $k$ is a normalization constant to be determined.
Under the modular transformation $t \rightarrow 1/t$ (which maps the closed 
string tree channel to the open string loop channel) we obtain the annulus 
amplitude ($G^{ij}={\tilde e}^i\cdot{\tilde e}^j$):
\begin{equation}
{\cal A} \sim {1\over 2^b} k \sum_{\bf{m}} e^{-(4 \pi t) G^{ij} m_i m_j}~,
\end{equation}
where $m_j\in {\bf Z}$ if $B_{ij}=0$, and $m_j\in {1\over 2}{\bf Z}$ if 
$B_{ij}=1/2$.
The annulus amplitude must be properly normalized 
so that the overall normalization factor has the interpretation 
of the number of D9-branes. That is, we must have 
$k/2^b=({\mbox{Tr}}(\gamma_{0,9}))^2$ (recall that 
${\mbox{Tr}}(\gamma_{0,9})=n_9$ is the total number of D9-branes). 
This implies that the cylinder amplitude (\ref{cylinder}), as well as the 
corresponding tadpole, is proportional to
$2^b({\mbox{Tr}}(\gamma_{0,9}))^2$ (instead of 
$({\mbox{Tr}}(\gamma_{0,9}))^2$ as in the case with $b=0$). 
Similar considerations for the M{\"o}bius amplitude
show that the corresponding tree channel amplitude (which is a cylinder
with closed strings propagating between a D9-brane and a cross-cap)
is proportional to $2^{b/2} {\mbox{Tr}}(\gamma_{0,9})$. The Klein bottle
contribution into the tadpoles for the untwisted R-R 10-form potential 
is not modified  as the orientifold projection keeps the left-right symmetric 
states ($P_L=P_R$) which are insensitive to the presence of the $B$-field
as can be seen from (\ref{momenta}). Putting all of the above together we
conclude that the tadpole cancellation condition 
for the untwisted R-R 10-form potential becomes:
\begin{equation}
2^{b} ({\mbox {Tr}} (\gamma_{0,9}))^2 - 2^{b/2} 64 
{\mbox {Tr}} (\gamma_{0,9}) + 32^2 = 0~.
\end{equation}
Therefore, the number of D9-branes is given by $n_9=32/2^{b/2}$.

{}Next, let us see what happens to the tadpoles for the untwisted R-R 6-form
potential (which determine the number of D5-branes) in the cases where the
orbifold group ${\bf Z}_N$ has even order $N\in 2{\bf N}$ (for odd $N$ there 
are no D5-branes present). The open strings stretched between D5-branes
satisfy Dirichlet boundary conditions. This implies that in the tree channel the 
closed strings must satisfy the condition of ``no winding flow'' through the 
boundaries (that is, D5-branes):
\begin{equation}
P_L - P_R = 0
\end{equation}
This does not impose any new constraints on the 55 cylinder amplitude 
in the presence of the $B$-field. However, the Klein bottle contribution into the
tadpoles for the untwisted R-R 6-form potential is modified as the corresponding
projection (that is, $\Omega R$ where $R$ is the generator of the ${\bf Z}_2$
subgroup of ${\bf Z}_N$) keeps the states with $P_L=-P_R$. 
The net result of this is that the Klein bottle tadpole is reduced by $2^{b}$.
The M{\"o}bius strip amplitude can be analyzed similarly, and the
tadpole cancellation condition
for the untwisted R-R 6-form potential becomes:
\begin{equation}
({\mbox {Tr}} (\gamma_{0,5}))^2 - {64 \over 2^{b/2}} {\mbox {Tr}} (\gamma_{0,5}) 
+ {1\over 2^b} 32^2 = 0 ~.
\end{equation}
Therefore, the number of D5-branes is given by $n_5=32/2^{b/2}$. 
This was also expected from T-duality between D9- and D5-branes.

{}So far we have focused on the $99$ and $55$ open string 
sectors. Let us understand the effect of the $B$-field on the 
$59$ open string sector. From the 99 and 55 cylinder amplitudes 
one can construct the boundary states corresponding to D9- and D5-branes
($p=9,5$):
\begin{equation}
{\cal C}_{pp} = \langle B,p \vert \exp \left( - \pi t (L_0 + \overline{L}_0) \right)
           \vert B,p \rangle ~.
\end{equation}
Let $\vert B,p \rangle_0$ be the corresponding boundary states without the
$B$-field. Then from the previous discussion it should be clear that
in the presence of the $B$-field the boundary states are given by
$\vert B,9 \rangle=2^{b/2} \vert B,9 \rangle_0$ and 
$\vert B,5 \rangle=\vert B,5 \rangle_0$.
The $59$ cylinder amplitude therefore reads:
\begin{equation}
{\cal C}_{59} = \langle B,5 \vert \exp \left( - \pi t (L_0 + \overline{L}_0) 
 \right)  \vert B,9 \rangle=2^{b/2} {\cal C}_{59}^{(0)} ~,
\end{equation}
where ${\cal C}_{59}^{(0)}$ is the 59 cylinder amplitude without the $B$-field.
The interpretation of the extra factor of $2^{b/2}$ in the 59 loop channel amplitude
is that
the $59$ open string sector states come with a multiplicity $\xi=2^{b/2}$
in the presence of the $B$-field. (Recall that without the $B$-field the 
multiplicity of states
in the 59 open string sector was one per given configuration of 
Chan-Paton charges \cite{GP,6Dorientifold}.)

\section{Orientifolds with Non-zero $B$-field}

{}In this section we discuss orientifolds of Type IIB on $T^4/{\bf Z}_N$
($N=2,3,4,6$) in the presence of the $B$-field. For illustrative purposes 
we will focus our discussion on the ${\bf Z}_2$ case, and state the results
for the other cases. 

{}First consider the ${\bf Z}_2$ orientifold model of 
\cite{PS,GP} without the $B$-field. 
The closed string sector gives 20 neutral
hypermultiplets (4 from the untwisted sector and 1 from each of the 16 fixed 
points in the twisted sector)
and 1 tensor multiplet (from the untwisted sector).
The open string sector gives gauge bosons and charged hypermultiplets.
The $99$ gauge group is $U(16)$ in the absence of Wilson lines,
and the $55$ gauge group is $U(16)$ when all the D5-branes are located at the 
same fixed point. The spectrum of this model is given in Table \ref{spectrum1}.
The states charged under both $99$ and $55$ gauge 
groups correspond to the $59$ open string sector. 

{}Let us turn on the $B$-field with $b=2$. From our discussions in section 
\ref{ReducedRank}, numbers of both the D9- and D5-branes are reduced to
16. The twisted tadpole cancellation conditions remain the same in
the presence of the $B$-field, and are given by \cite{GP}:
\begin{equation}
{\mbox {Tr}} (\gamma_{1,9}) ={\mbox {Tr}} (\gamma_{1,5})=0~.
\end{equation} 
Using the solutions for the Chan-Paton gamma matrices, we can find the
open string spectrum. The gauge group is reduced to 
$U(8)_{99} \otimes U(8)_{55}$.
The massless spectrum of this model is summarized in 
Table \ref{spectrum1}. The multiplicity $\xi=2$ in the 59 sector (that is, presence 
of {\em two} (instead of one as in the previous case) bi-fundamentals 
$({\bf 8};{\bf 8})$ of $U(8)_{99} \otimes U(8)_{55}$) 
is related to the
corresponding multiplicity in the 
$59$ cylinder amplitude as we discussed in section II. 

{}Before we discuss the closed string spectrum, let us see what we
would expect from anomaly cancellation. Since we are 
compactifying Type IIB 
on an orbifold limit of K3 (which has $80$ moduli), the number
of closed string 
hypermultiplets plus extra tensor multiplets must be $20$:
\begin{equation}
n_H^{c} + n_T = 20~.
\end{equation} 
The anomaly cancellation condition becomes:
\begin{equation}
n_H^{o} - n_V = 224 - 28 n_T~,
\end{equation}
where $n_H^o$ is the number of open string hypermultiplets. 
In the ${\bf Z}_2$ model with $b=2$, the above condition implies 
that $n_T=4$. Therefore, in addition to rank reduction, some of
the closed string hypermultiplets must be converted to tensor multiplets
when we turn on the $B$-field.

{}To see that there are precisely $4$ extra tensor multiplets when we
turn on the $B$-field with $b=2$, let us analyze the closed string spectrum
more carefully. Recall that in the case without the $B$-field, the untwisted
sector gives 4 hypermultiplets and one tensor multiplet.
At each fixed point of the twisted sector, the NS and R sector 
massless states transform in the following representations of the
six dimensional little group $SO(4)\approx SU(2) \otimes SU(2)$:
\begin{equation}
\begin{array}{lc}
{\mbox {Sector}} \quad \quad & SU(2) \otimes SU(2) ~ {\mbox {rep.}} \\
{\mbox {NS}}     & 2 ({\bf 1},{\bf 1})~, \\
{\mbox {R}}      & ({\bf 1},{\bf 2})~. \\
\end{array}
\end{equation}
The twisted sector spectrum is obtained by taking products
of states from the left- and right-moving sectors. The orientifold
projection $\Omega$ keeps symmetric combinations in the NS-NS sector
and antisymmetric combinations in the R-R sector. This gives 
$4({\bf 1},{\bf 1})$ which is the bosonic content of a hypermultiplet.
Since there are $16$ fixed points in the ${\bf Z}_2$ orbifold,
the twisted sector gives total of $16$ hypermultiplets.

{}However, the above counting is based on the fact that the orbifold 
fixed points
are invariant under $\Omega$. If a fixed point picks up a minus sign 
under the  $\Omega$ projection ({\it i.e.}, if 
it is odd under the $\Omega$
projection), then
the states that are kept after the $\Omega$ projection 
are antisymmetric combinations in the NS-NS sector and
symmetric combinations in the R-R sector. This gives rise to 
$({\bf 1},{\bf 1}) \oplus ({\bf 1},{\bf 3})$ which is the bosonic
content of a tensor multiplet.

{}Therefore, we expect that four of the ${\bf Z}_2$ fixed points 
are not $\Omega$ invariant for $b=2$. To see that this is indeed the case, 
let us go to 
the enhanced symmetry point such that the momentum lattice
$\{(P_L,P_R)\}\equiv \Gamma^{4,4}$ is the $SO(8)$ lattice, {\it i.e.}, 
$P_L,P_R \in \tilde{\Gamma}^4$ ($SO(8)$ weight lattice), and
$P_L-P_R \in \Gamma^4$ ($SO(8)$ root lattice). This is achieved by
choosing specific values of $B_{ij}$ and $G_{ij}$:
\begin{equation}
	B_{ij}=\left( \begin{array}{cccc}
               0 & 1/2 & 0 & -1/2\\
               -1/2 & 0 & 1/2 & 0\\
               0 & -1/2 & 0 & 1/2 \\
               1/2 & 0 & -1/2 & 0
               \end{array}
        \right)~,~~~
 G_{ij} =\left( \begin{array}{cccc}
               2 & -1 & 0 & 1\\
               -1 & 2 & -1 & 0\\
               0 & -1 & 2 & -1\\
               1 & 0 & -1 & 2
               \end{array}
        \right)~.
\end{equation}
Note that the NS-NS antisymmetric tensor $B_{ij}$ has rank $b=2$ in this case.

{}At this special point with $SO(8)$ enhanced symmetry, we can 
express the $16$ ${\bf Z}_2$ fixed points as points in the
$\Gamma^{4,4}$ lattice. If we write the $SO(8)$ representations
in the $SU(2)^4$ basis, the fixed points are given by 
$P_L = P_R = ({\bf 2},{\bf 1},{\bf 1},{\bf 1})$, 
$({\bf 1},{\bf 2},{\bf 1},{\bf 1})$,
$({\bf 1},{\bf 1},{\bf 2},{\bf 1})$,
and $({\bf 1},{\bf 1},{\bf 1},{\bf 2})$.

{}Let us focus on the four fixed points with
$P_L=P_R=({\bf 2},{\bf 1},{\bf 1},{\bf 1})$ as the analysis for
the other fixed points is similar.
We can form three linear combinations of these four fixed points such that
they are invariant under the $\Omega$ projection ({\it i.e.}, 
$P_L \leftrightarrow P_R$). The remaining linear combination   
is not $\Omega$ invariant but has $\Omega=-1$.
Therefore, there are altogether $12$ fixed points with $\Omega=+1$
(which give 12 hypermultiplets) 
and $4$ fixed points with $\Omega=-1$ (which give $4$ tensor
multiplets).

{}Although our arguments are made at the enhanced symmetry point,
the results hold for generic points as well. This is because
rank reduction of the gauge group depends only on the rank
of $B_{ij}$ and not on its precise form nor the
values of $G_{ij}$. We can reach other points in the moduli space
with different
$\Gamma^{4,4}$ lattices by changing $G_{ij}$. In particular, the number of tensor 
multiplets in the ${\bf Z}_2$ twisted sector is always 4 as long as $b=2$.

{}Since we are compactifying Type IIB on $T^4/{\bf Z}_2$, we can consider 
turning on $B_{ij}$ with $b=4$. We can immediately deduce that
the number of extra tensor multiplets in this case is $n_T =6$.
To see this, first consider a compactification on $(T^2\otimes T^2)/{\bf Z}_2$.
Let the $B$-field in the first 2-torus be
non-zero, while keeping the $B$-field in 
the second 2-torus zero. The fixed points decompose as ${\bf 16}= {\bf 4}
\otimes {\bf 4}$,
where the second factor of ${\bf 4}$ is $\Omega$ invariant as there is no $B$-field 
in the second 2-torus. But from the above analysis we know that the total number
of fixed points for $b=2$ (which is the case here) must be 12. This implies that
the first factor of ${\bf 4}$ (coming from the first 2-torus with the $B$-field) 
decomposes
as ${\bf 4}={\bf 3}_+\oplus{\bf 1}_-$, where the subscript indicates whether the 
corresponding fixed points are even or odd under $\Omega$. 
Now consider compactification on $(T^2\otimes T^2)/{\bf Z}_2$ with non-zero 
$B$-field in both 2-tori. This corresponds to $b=4$.
The fixed points now decompose into
${\bf 16}= ( {\bf 3}_+ \oplus{\bf 1}_- ) \otimes ({\bf 3}_+ \oplus {\bf 1}_-) = 
{\bf 10}_+ \oplus{\bf 6}_-$. 
The number of D9-branes as well as the number of D5-branes
in this case is 8. The maximal unbroken
gauge group is $U(4) \otimes U(4)$. The massless spectrum of this model is given 
in Table \ref{spectrum1}. Note that the anomaly is cancelled in this model.

{}Next, we briefly discuss other ${\bf Z}_N$ orientifolds.
Note that the only sector where the number of 
tensor multiplets can vary with the $B$-field is the ${\bf Z}_2$ twisted sector.
This is because the action of $\Omega$ interchanges the $g^k$ twisted sector
with the $g^{N-k}$ twisted sector for $2k\not=0,N$ \cite{6Dorientifold}. 
Each of the fixed points in such sectors, therefore, gives rise to one hypermultiplet
and one tensor multiplet regardless of the $B$-field. Thus, in the ${\bf Z}_3$ case 
the number of extra tensor multiplets is always $n_T=9$.
In the ${\bf Z}_4$ case in the ${\bf Z}_2$ twisted sector 4 (of the original 
16) fixed points are also fixed under ${\bf Z}_4$ ({\em i.e.}, they are invariant under
${\bf Z}_4$). The other 12 fixed points pair up into 6 ${\bf Z}_4$ invariant pairs. 
For $b=2$, two of these 6 pairs  
have $\Omega=-1$ (which give
2 tensor multiplets).
Therefore, the total number of extra tensor multiplets in this case is $n_T=6$.
For $b=2$, three of the 6 pairs  
have $\Omega=-1$ (which give
3 tensor multiplets), and 
the total number of extra tensor multiplets in this case is $n_T=7$.
In the ${\bf Z}_6$ case 1 (of the original 
16) fixed point in the ${\bf Z}_2$ twisted sector
is also fixed under ${\bf Z}_3$ ({\em i.e.}, it is invariant under
${\bf Z}_3$). The other 15 fixed points form 5 linear combinations invariant under
${\bf Z}_3$, 5 linear combinations that pick up a phase $\omega=\exp(2\pi i/3)$
under ${\bf Z}_3$, and 5 linear combinations that pick up a phase
$\omega^2$ under ${\bf Z}_3$. The $\Omega=-1$ states are always among the
10 combinations which are not invariant under ${\bf Z}_3$. That is, all the
${\bf Z}_2$ fixed points which are not invariant under $\Omega$ in the presence 
of the $B$-field do not contribute to the massless spectrum. Thus, the number 
of extra tensor multiplets in the ${\bf Z}_6$ model is always $n_T=6$ regardless 
of the value of $b$.

{}The massless open and closed string
spectra of the ${\bf Z}_N$ orientifolds with and without the $B$-field are given
in Table \ref{spectrum1} (for $N=2,4$) and Table \ref{spectrum2} (for $N=3,6$).

\section{Comments}

{}In this section we point out relations between various orientifolds with and 
without the $B$-field, and also discuss the F-theory duals of these models.
For convenience we will refer to the ${\bf Z}_N$ model with the rank-$b$
$B$-field as the $[N,b]$ model. (Here we note that the $[6,4]$ model was discussed in
the third reference in \cite{PS}, while the $[3,4]$ model was discussed in \cite{Ang}.) 

{}To begin with note the following.\\
$\bullet$ The $[2,2]$ and $[4,0]$ models are on the same moduli. This can be 
seen by starting from the  $[4,0]$ model and giving vevs to the fields
$({\bf 8},{\bf 8};{\bf 1},{\bf 1})$ and $({\bf 1},{\bf 1};{\bf 8},{\bf 8})$ such that
the 99 gauge group is broken to $[U(8)_{\mbox{\small{diag}}}]_{99}\subset
[U(8)\otimes U(8)]_{99}$, and similarly for the 55 gauge group.\\
$\bullet$ The $[2,4]$, $[4,2]$, $[6,0]$, $[6,2]$ and $[6,4]$ models are on the same 
moduli. This is obvious for the $[4,2]$ and $[6,2]$ models as their massless 
spectra are identical. Similarly, the massless spectra of the $[2,4]$ and $[6,4]$
models are the same. We can obtain the $[2,4]$ model from the $[4,2]$ model
by giving vevs to the fields
$({\bf 4},{\bf 4};{\bf 1},{\bf 1})$ and $({\bf 1},{\bf 1};{\bf 4},{\bf 4})$ such that
the 99 gauge group is broken to $[U(4)_{\mbox{\small{diag}}}]_{99}\subset
[U(4)\otimes U(4)]_{99}$, and similarly for the 55 gauge group. In fact, we can 
Higgs the gauge group completely. After Higgsing there are 56 neutral 
hypermultiplets in the open string sector. Note that in the $[6,0]$ model the 
gauge group can also be Higgsed completely, and are 56 neutral
hypermultiplets in the open string sector in this case as well. This implies that
the $[6,0]$ model is on the same moduli as the other four models.\\
$\bullet$ The $[3,2]$ and $[3,4]$ models are on the same moduli. This can be seen
by starting from the $[3,2]$ model and giving a vev to the field in the symmetric 
(${\bf 36}$) representation of $U(8)$ such that the 99 gauge group is broken to
$SO(8)$. 

{}Here we can extend some of the analyses of \cite{gj} to the orientifolds 
with non-zero $B$-field. In particular, let us compactify these models further
on $T^2$. The resulting four dimensional models have ${\cal N}=2$ 
space-time supersymmetry. Let us go to a generic point in the moduli space
where the gauge group is maximally Higgsed (so that the gauge group
is either completely broken or consists of Abelian factors only). Let $r(V)$ be the
number of open string sector vector multiplets after Higgsing. Then the total
number of vector multiplets is given by $r(V)+T+2$, where $T=n_T+1$ is the 
total number of tensor multiplets in six dimensions. Let $H^0$ be the number of
neutral hypermultiplets that are uncharged with respect to the left-over 
Abelian gauge group. Then, if we assume that the four dimensional models 
have Type IIA duals, the Hodge numbers of the corresponding Calabi-Yau
three-folds are given by \cite{vafa}:
\begin{eqnarray}
h^{1,1}&=&r(V)+T+2 ~,\\
h^{2,1}&=&H^{0} -1 ~.
\end{eqnarray}
{}From the various dualities between Type IIA, Heterotic, Type I and F-theory,
it is not difficult to see that these Calabi-Yau three-folds must be elliptically
fibered, and F-theory compactifications on these spaces should be dual to 
the original six dimensional orientifold models. In the table below we give the Hodge
(and Euler) numbers
of the Calabi-Yau three-folds for each of the $[N,b]$ models.

\begin{center}
\begin{tabular}{|c|c|c|c|c|} \hline
\phantom{***} Model \phantom{***}
& \phantom{***} $b$ \phantom{***}
& \phantom{***} Gauge Group \phantom{***}
& \phantom{***} $(h^{1,1},h^{2,1})$ \phantom{***}
& \phantom{***} $\chi$ \phantom{***} \\ \hline
${\bf Z}_2$ & 0 & $U(16) \otimes U(16)$ 
            & $(3,243)$ & $-480$ \\ \hline
            & 2 & $U(8) \otimes U(8)$  
            & $(7,127)$ & $-240$ \\ \hline
            & 4 & $U(4) \otimes U(4)$ 
            & $(9,69)$ & $-120$ \\ \hline
${\bf Z}_3$ & 0 & $U(8) \otimes SO(16)$ 
            & $(20,14)$ & $12$ \\ \hline
            & 2 & $U(8)$  
            & $(16,10)$ & $12$ \\ \hline
            & 4 & $SO(8)$ 
            & $(16,10)$ & $12$ \\ \hline
${\bf Z}_4$ & 0 & $[U(8) \otimes U(8)]^2$ 
            & $(7,127)$ & $-240$ \\ \hline
            & 2 & $[U(4) \otimes U(4)]^2$ 
            & $(9,69)$ & $-120$ \\ \hline
            & 4 & $[U(2) \otimes U(2)]^2$
            & $(10,40)$ & $-60$ \\ \hline
${\bf Z}_6$ & 0 & $[U(4) \otimes U(4) \otimes U(8)]^2$ 
            & $(9,69)$ & $-120$ \\ \hline
            & 2 & $[U(4) \otimes U(4) ]^2$
            & $(9,69)$ & $-120$ \\ \hline
            & 4 & $U(4) \otimes U(4)$
            & $(9,69)$ & $-120$ \\ \hline
\end{tabular}
\end{center}

\acknowledgements

{}We would like to thank Piljin Yi for discussions. 
The research of G.S. and S.-H.H.T. was partially
supported by the 
National Science Foundation. G.S. would also like to thank
Joyce M. Kuok Foundation for financial support.
The work of Z.K. was supported in part by the grant NSF PHY-96-02074, 
and the DOE 1994 OJI award. 
Z.K. would also like to thank Albert and Ribena Yu for 
financial support.

%%%%%%%%%%%%%Table I %%%%%%%%
%%%%%%%%%%%%%%%%%%%%%%%%%%%%%%%%%%%%%%%%%%%%%%%%%%%%%%%%%%%%%%%%%%%%%%%%%%%%%%%
\begin{table}[t]
\begin{tabular}{|c|c|c|l|c|c|}
%%%%%%%%%%%%%%%%%%%%%%%%%%%%%%%%%%%%%%%%%%%%%%%%%%%%%%%%%%%%%%%%%%%%%%%%%%%%
 Model & $b$ & Gauge Group & \phantom{Hy} Charged  & Neutral 
& Extra Tensor  \\
       &   &             &Hypermultiplets & Hypermultiplets
&Multiplets \\
\hline
%%%%%%%%%%%%%%%%%%%%%%%%%%%%%%%%%%%%%%%%%%%%%%%%%%%%%%%%%%%%%%%%%%%%%%%%%%%
${\bf Z}_2$ & 0 & $U(16)_{99} \otimes U(16)_{55}$ & 
 $2 ({\bf 120};{\bf 1})$ & $20$
& $0$ \\
            &   &                       & $2 ({\bf 1};{\bf 120})$ & & \\
            &   &                       & $({\bf 16};{\bf 16})$ & & \\
\hline
${\bf Z}_2$ & 2 & $U(8)_{99} \otimes U(8)_{55}$ & $2 ({\bf 28};{\bf 1})$ & $16$
& $4$ \\
            &   &                       & $2 ({\bf 1};{\bf 28})$ & & \\
            &   &                       & $2 ({\bf 8};{\bf 8})$ & & \\
\hline
${\bf Z}_2$ & 4 & $U(4)_{99} \otimes U(4)_{55}$ & $2 ({\bf 6};{\bf 1})$ & $14$
& $6$ \\
            &   &                       & $2 ({\bf 1};{\bf 6})$ & & \\
            &   &                       & $4 ({\bf 4};{\bf 4})$ & & \\
\hline
%%%%%%%%%%%%%%%%%%%%%%%%%%%%%%%%%%%%%%%%%%%%%%%%%%%%%%%%%%%%%%%%%%%%%%%%
${\bf Z}_4$ & 0 & $[U(8) \otimes U(8)]_{99}\otimes$ 
& $({\bf 28},{\bf 1};{\bf 1},{\bf 1})$ & $16$ & $4$ \\ 
& &$[U(8) \otimes U(8)]_{55}$ & $({\bf 1},{\bf 28};{\bf 1},{\bf 1})$ & & \\
& & & $({\bf 8},{\bf 8};{\bf 1},{\bf 1})$ & & \\
& & & $({\bf 1},{\bf 1};{\bf 28},{\bf 1})$ & & \\
& & & $({\bf 1},{\bf 1};{\bf 1},{\bf 28})$ & & \\
& & & $({\bf 1},{\bf 1};{\bf 8},{\bf 8})$ & & \\
& & & $({\bf 8},{\bf 1};{\bf 8},{\bf 1})$ & & \\
& & & $({\bf 1},{\bf 8};{\bf 1},{\bf 8})$ & & \\
\hline
%%%%%%%%%%%%%%%%%%%%%%%%%%%%%%%%%%%%%%%%%%%%%%%%%%%%%%%%%%%%%%%%%%%%%%%%%
${\bf Z}_4$ & 2 & $[U(4) \otimes U(4)]_{99}\otimes$ 
& $({\bf 6},{\bf 1};{\bf 1},{\bf 1})$ & $14$ & $6$ \\
& & $[U(4) \otimes U(4)]_{55}$& $({\bf 1},{\bf 6};{\bf 1},{\bf 1})$ & & \\
& & & $({\bf 4},{\bf 4};{\bf 1},{\bf 1})$ & & \\
& & & $({\bf 1},{\bf 1};{\bf 6},{\bf 1})$ & & \\
& & & $({\bf 1},{\bf 1};{\bf 1},{\bf 6})$ & & \\
& & & $({\bf 1},{\bf 1};{\bf 4},{\bf 4})$ & & \\
& & & $2 ({\bf 4},{\bf 1};{\bf 4},{\bf 1})$ & & \\
& & & $2 ({\bf 1},{\bf 4};{\bf 1},{\bf 4})$ & & \\
\hline
%%%%%%%%%%%%%%%%%%%%%%%%%%%%%%%%%%%%%%%%%%%%%%%%%%%%%%%%%%%%%%%%%%%%%%%%%
${\bf Z}_4$ & 4 & $[U(2) \otimes U(2)]_{99}\otimes$ 
& $4 ({\bf 1},{\bf 1};{\bf 1},{\bf 1})$ & $13$ & $7$ \\
& & $[U(2) \otimes U(2)]_{55}$& $({\bf 2},{\bf 2};{\bf 1},{\bf 1})$ & & \\
& & & $({\bf 1},{\bf 1};{\bf 2},{\bf 2})$ & & \\
& & & $4 ({\bf 2},{\bf 1};{\bf 2},{\bf 1})$ & & \\
& & & $4 ({\bf 1},{\bf 2};{\bf 1},{\bf 2})$ & & \\
\hline
%%%%%%%%%%%%%%%%%%%%%%%%%%%%%%%%%%%%%%%%%%%%%%%%%%%%%%%%%%%%%%%%%%%%%%%%%%
\end{tabular}
%%%%%%%%%%%%%%%%%%%%%%%%%%%%%%%%%%%%%%%%%%%%%%%%%%%%%%%%%%%%%%%%%%%%%%%%%%%
\caption{The massless spectrum of the six dimensional Type IIB orientifolds
on $T^4/{\bf Z}_N$ for $N=2,4$, and various values of $b$ (the rank of $B_{ij}$).
The semi-colon in the column ``Charged Hypermultiplets'' separates $99$ and 
$55$ representations.}
\label{spectrum1} 
\end{table}
%%%%%%%%%%%%%%%%%%%%%%%%%%%%%%%%%%%%%%%%%%%%%%%%%%%%%%%%%%%%%%%%%%%%%%%%%%%%%%%

%%%%%%%%%%%%%Table II %%%%%%%%
%%%%%%%%%%%%%%%%%%%%%%%%%%%%%%%%%%%%%%%%%%%%%%%%%%%%%%%%%%%%%%%%%%%%%%%%%%%%%%%
\begin{table}[t]
\begin{tabular}{|c|c|c|l|c|c|}
%%%%%%%%%%%%%%%%%%%%%%%%%%%%%%%%%%%%%%%%%%%%%%%%%%%%%%%%%%%%%%%%%%%%%%%%%%%%
Model & $b$ & Gauge Group & \phantom{Hy} Charged  & Neutral 
& Extra Tensor  \\
       &   &             &Hypermultiplets & Hypermultiplets
&Multiplets \\
\hline
%%%%%%%%%%%%%%%%%%%%%%%%%%%%%%%%%%%%%%%%%%%%%%%%%%%%%%%%%%%%%%%%%%%%%%%%%%%
${\bf Z}_3$ & 0 & $[U(8) \otimes SO(16)]_{99}$ & $({\bf 28},{\bf 1})$ & $11$
& $9$ \\
& & & $({\bf 8},{\bf 16})$ & & \\
\hline
& 2 & $U(8)_{99}$ & ${\bf 36}$ & $11$
& $9$ \\
\hline
& 4 & $SO(8)_{99}$ & --- & $11$
& $9$ \\
\hline
%%%%%%%%%%%%%%%%%%%%%%%%%%%%%%%%%%%%%%%%%%%%%%%%%%%%%%%%%%%%%%%%%%%%%%%%
${\bf Z}_6$ & 0 & $[U(4) \otimes U(4) \otimes U(8)]_{99}\otimes$ 
&$({\bf 6},{\bf 1},{\bf 1};{\bf 1},{\bf 1},{\bf 1})$ & $14$ & $6$ \\
& & $[U(4) \otimes U(4) \otimes U(8)]_{55}$&
$({\bf 1},{\bf 6},{\bf 1};{\bf 1},{\bf 1},{\bf 1})$  & & \\
& & & $({\bf 4},{\bf 1},{\bf 8};{\bf 1},{\bf 1},{\bf 1})$ & & \\
& & & $({\bf 1},{\bf 4},{\bf 8};{\bf 1},{\bf 1},{\bf 1})$ & & \\
& & & $({\bf 1},{\bf 1},{\bf 1};{\bf 6},{\bf 1},{\bf 1})$ & & \\
& & & $({\bf 1},{\bf 1},{\bf 1};{\bf 1},{\bf 6},{\bf 1})$ & & \\
& & & $({\bf 1},{\bf 1},{\bf 1};{\bf 4},{\bf 1},{\bf 8})$ & & \\
& & & $({\bf 1},{\bf 1},{\bf 1};{\bf 1},{\bf 4},{\bf 8})$ & & \\
& & & $({\bf 4},{\bf 1},{\bf 1};{\bf 4},{\bf 1},{\bf 1})$ & & \\
& & & $({\bf 1},{\bf 4},{\bf 1};{\bf 1},{\bf 4},{\bf 1})$ & & \\
& & & $({\bf 1},{\bf 1},{\bf 8};{\bf 1},{\bf 1},{\bf 8})$ & & \\
\hline
%%%%%%%%%%%%%%%%%%%%%%%%%%%%%%%%%%%%%%%%%%%%%%%%%%%%%%%%%%%%%%%%%%%%%%%%
& 2 & $[U(4) \otimes U(4)]_{99}\otimes$ 
& $({\bf 6},{\bf 1};{\bf 1},{\bf 1})$ & $14$ & $6$ \\
& &$[U(4) \otimes U(4)]_{55}$& $({\bf 1},{\bf 6};{\bf 1},{\bf 1})$ & & \\
& & & $({\bf 4},{\bf 4};{\bf 1},{\bf 1})$
& & \\
& & & $({\bf 1},{\bf 1};{\bf 6},{\bf 1})$
& & \\
& & & $({\bf 1},{\bf 1};{\bf 1},{\bf 6})$
& & \\
& & & $({\bf 1},{\bf 1};{\bf 4},{\bf 4})$
& & \\
%%%%%%%%%%%%%%%%%%%%%%%%%%%%%%%%%%%%%%%%%%%%%%%%%%%%%%%%%%%%%%%%%%%%%%%%%%%
& & & $2 ({\bf 4},{\bf 1};{\bf 4},{\bf 1})$
& & \\
& & & $2 ({\bf 1},{\bf 4};{\bf 1},{\bf 4})$
& & \\
\hline
%%%%%%%%%%%%%%%%%%%%%%%%%%%%%%%%%%%%%%%%%%%%%%%%%%%%%%%%%%%%%%%%%%%%%%%%%%%
& $4$ & $U(4)_{99} \otimes U(4)_{55}$ 
& $2({\bf 6};{\bf 1})$ & $14$ & $6$ \\
& & & $2({\bf 1};{\bf 6})$ & & \\
& & & $4({\bf 4};{\bf 4})$ & & \\
\end{tabular}
%%%%%%%%%%%%%%%%%%%%%%%%%%%%%%%%%%%%%%%%%%%%%%%%%%%%%%%%%%%%%%%%%%%%%%%%%%%
\caption{The massless spectrum of the six dimensional Type IIB orientifolds
on $T^4/{\bf Z}_N$ for $N=3,6$, and various values of $b$ (the rank of $B_{ij}$).
The semi-colon in the column ``Charged Hypermultiplets'' separates $99$ and 
$55$ representations.}
\label{spectrum2} 
\end{table}
%%%%%%%%%%%%%%%%%%%%%%%%%%%%%%%%%%%%%%%%%%%%%%%%%%%%%%%%%%%%%%%%%%%%%%%%%%%%%%%


\begin{references}

\bibitem{PS} G. Pradisi and A. Sagnotti, Phys. Lett. {\bf B216} (1989) 59;\\
M. Bianchi and A. Sagnotti, Phys. Lett. {\bf B247} (1990) 517; Nucl. Phys. 
{\bf B361} (1991) 539. 

\bibitem{GP} E.G. Gimon and J. Polchinski, Phys. Rev. {\bf D54} (1996) 1667, 
hep-th/9601038.

\bibitem{6Dorientifold}
E.G. Gimon and C.V. Johnson, Nucl. Phys. {\bf B477} (1996) 715, 
hep-th/9604129;\\
A. Dabholkar and J. Park, Nucl. Phys. {\bf B477} (1996) 701, 
hep-th/9604178.



\bibitem{6Dduality} 
A. Dabholkar and J. Park, Nucl. Phys. {\bf B472} (1996) 207, 
hep-th/9602030; Phys. Lett. {\bf B394} (1997) 302, 
hep-th/9607041;\\
J. Polchinski, Phys. Rev. {\bf D55} (1997) 6423, hep-th/9606165;\\
J.D. Blum and A. Zaffaroni, Phys. Lett. {\bf B387} (1996) 71, 
hep-th/9607019;\\
J.D. Blum, Nucl. Phys. {\bf B486} (1997) 34, hep-th/9608053.

\bibitem{BL} M. Berkooz and R.G. Leigh, Nucl. Phys. {\bf B483} (1997) 187,
hep-th/9605049.

\bibitem{Sagnotti} C. Anjelantonj, M. Bianchi, G. Pradisi, A. Sagnotti and 
Ya.S. Stanev, Phys. Lett. {\bf B385} (1996) 96, hep-th/9606169.

\bibitem{ZK} Z. Kakushadze, Nucl. Phys. {\bf B512} (1998) 221, hep-th/9704059.

\bibitem{KS} Z. Kakushadze and G. Shiu, Phys. Rev. {\bf D56} (1997) 3686,
hep-th/9705163; Nucl. Phys. {\bf 520} (1998) 75, hep-th/9706051.

\bibitem{Zw} G. Zwart, hep-th/9708040.

\bibitem{KST} Z. Kakushadze, G. Shiu and S.-H.H. Tye, Nucl. Phys. {\bf B533} (1998) 25, hep-th/9804092.

\bibitem{PW} J. Polchinski and E. Witten, Nucl. Phys. {\bf B460} (1996) 
525, hep-th/9510169

\bibitem{Bij} 
M. Bianchi, G. Pradisi and A. Sagnotti, Nucl. Phys. {\bf B376} (1992) 365.

\bibitem{bianchi}
M. Bianchi, hep-th/9711201.

\bibitem{toroidal}
E. Witten, hep-th/9712028. 

\bibitem{anomalies}
M.B. Green, J.H. Schwarz and P.C. West, Nucl. Phys. {\bf B254} (1985) 327;\\
J. Erler, J. Math. Phys. {\bf 35} (1994) 1819.

\bibitem{SS} A. Sen and S. Sethi, Nucl. Phys. {\bf B499} (1997) 45, 
hep-th/9703157.

\bibitem{Ang} C. Angelantonj, M. Bianchi, G. Pradisi, A. Sagnotti and Ya.S. Stanev,
Phys. Lett. {\bf B387} (1996) 743.

\bibitem{gj}E.G. Gimon and C.V. Johnson, Nucl. Phys. {\bf B479} (1996) 285, 
hep-th/9606176.

\bibitem{vafa} C. Vafa, Nucl. Phys. {\bf B469} (1996) 403, hep-th/9602022;\\
D.R. Morrison and C. Vafa, Nucl. Phys. {\bf B473} (1996) 74, 
hep-th/9602114; Nucl. Phys. {\bf B476} (1996) 437, hep-th/9603161.

\end{references}
\end{document}